\documentclass[showpacs,preprintnumbers,prl,a4paper,twocolumn]{revtex4}
\usepackage{amssymb}
\usepackage{amsfonts}
\usepackage{amsmath}
\usepackage{graphicx}
\usepackage{dcolumn}
\usepackage{bm}

\begin{document}

\preprint{}
\title{\bf Dynamics and Synchrony from Oscillatory Data
 via Dimension Reduction}
\author{Jie Zhang$^1$, Kai Zhang$^2$, Jianfeng Feng$^3$, Junfeng Sun$^1$, Xiaoke Xu$^1$, Michael Small$^1$\\
\small{\emph{$^1$Department of Electronic and Information
Engineering, Hong Kong
Polytechnic University, Hong Kong\\
$^2$Department of Computer Science and Engineering, Hong Kong
University of Science  and Technology, Hong Kong\\ $^3$Department of
Computer Science and Mathematics, Warwick University, Coventry CV4
7AL, UK}} }

\email[$^1$]{enzhangjie@eie.polyu.edu.hk}
\date{\today }

\begin{abstract}
Complex, oscillatory data arises from a large variety of biological,
physical, and social systems. However, the inherent oscillation and
ubiquitous noise pose great challenges to current methodology such
as linear and nonlinear time series analysis. We exploit the state
of the art technology in pattern recognition and specifically,
dimensionality reduction techniques, and propose to rebuild the
dynamics accurately on the ``cycle'' scale. This is achieved by
deriving a compact representation of the cycles through global
optimization, which effectively preserves the topology of the cycles
that are embedded in a high dimensional Euclidian space. Our
approach demonstrates a clear success in capturing the intrinsic
dynamics and the subtle synchrony pattern from uni/bivariate
oscillatory data over traditional methods. Application to the human
locomotion data reveals important dynamical information which allows
for a clinically promising discrimination between healthy subjects
and those with neural pathology. Our results also provide
fundamental implications for understanding the neuromuscular control
of human walking.

\end{abstract}
\pacs{05.45.Tp, 05.45.Xt}
 \maketitle

Time series exhibiting complex oscillatory dynamics are widely
observed in the real world: from finance to biomedical systems
\cite{leo}. In this Letter we focus on an important class of data
with a stable oscillation frequency but irregular waveform
fluctuations, also called pseudoperiodic time series. Such data
arise from broad application domains and have gained particular
interest in recent years, with examples ranging from human ECG and
gait data \cite{ecg, Dingwell}, white blood-cell count and tremor
data in patients \cite{blood,tremor}, epidemic dynamics \cite{epi},
light intensity of laser \cite{laser}, sun spot numbers and global
temperature variation \cite{sun, temp}. Accurate extraction and
characterization of the dynamics of these time series is a general
problem, which holds the key to understanding the inner workings of
many important physical and biological systems. To this end,
traditional methods rely primarily on linear spectral analysis or
computing nonlinear characteristics, and recent attempts include
producing pseudoperiodic surrogate data \cite{pps}, or to establish
a nonlinear transform from cycles in time domain to nodes in a dual
complex network domain \cite{zj}.

However, there are still no generic, systematic and robust
approaches to handle such oscillatory time series. The Fourier
transform, though widely applied to oscillatory data, is inherently
a linear method and cannot account for the nonlinear nature of the
data, if any. For nonlinear approaches, the cyclic trend overlying
the signal, together with noise from unknown sources, can often mask
the intrinsic dynamics \cite{periodicity, noise} and pose great
challenges to the nonlinear time series analysis/modeling
techniques. For example, the most popular Poincare section method,
which reduces the flow data to intersections of the trajectories
with a lower-dimensional subspace, may produce highly corrupted
results under noisy environment, in that the intersection points can
no longer preserve the original dynamics in the presence of noise.
Things get even worse when the cycles possess complex noncoherent
waveforms. In this paper we propose a novel, generic approach that
can effectively capture the dynamics of oscillatory time series. By
mapping the cycles in the time series to a multi-dimensional
Euclidian space, we seek a low-dimensional representation which
topologically preserves the important proximity relation among
cycles, through efficient global optimization. It is the first time
that advances in dimension reduction are explored in reconstructing
the dynamics of complex oscillatory data. Our approach utilizes the
richer information of pairwise cycle correlation, therefore it not
only excavates the inherent dynamics obscured by the cyclic trend,
but also offers an extra robustness to noise due to the global
nature of the method. Exploration of our approach to probing the
synchrony between bivariate oscillatory time series is shown to be
able to reveal the subtle synchrony for which traditional approaches
will fail. We applied the proposed method to the human locomotion
data, and are capable of discriminating between the healthy subjects
and those with neuropathology reliably. Based on the results we are
able to make the important biological inference that the human
walking is not critically dependent on the peripheral neural
feedback.

\section{Theory}
\subsection{Reconstructing Dynamics Underlying Cyclic Trend by Spectral Clustering}
\begin{figure}
 \begin{center}
 \includegraphics[height=0.3\textwidth]{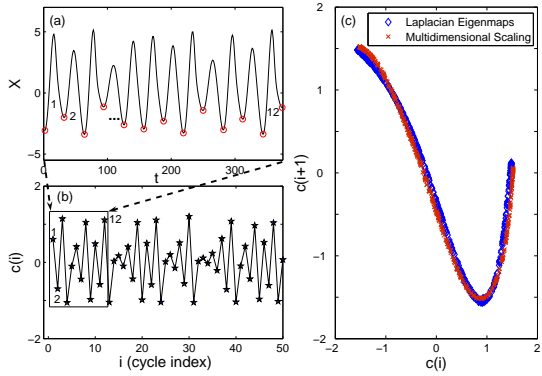}
 \caption{ (a) Time series form $X$-component of the chaotic R\"{o}ssler system,
which is divided into cycles by local minima (circles). (b)
Laplacian embedding $c(i)$ extracted on the cycle scale for the time
series shown in (a), and (c) the return plots of $c(i)$ derived by
two dimension reduction methods, i.e., the Laplacian Eigenmaps and
 Multidimentional scaling.} \label{ts}
\end{center}
 \end{figure}

We demonstrate how intrinsic dynamics of psuedoperiodic data can be
extracted using \emph{Laplacian Eigenmaps} \cite{laplacian}, a
nonlinear dimension reduction method based on \emph{spectral graph
theory}. We illustrate with the $X$-component of the chaotic
R\"{o}ssler system. Motivated by the fact that such data usually
exhibit a highly redundant patterns in the form of repeated cycles,
we first partition the time series into individual cycles $C_{i}$'s
($i=1,...,k$) by local minimums as shown in Fig. \ref{ts} (a). Each
cycle is then mapped to a high dimensional vector $x_i$ whose
dimension equals the number of points in that cycle. Our goal is to
compute a set of new, low-dimensional (in the simplest case, $1D$)
representation, or embedding $y_i$'s, such that the proximity
relations among $x_i$'s are maximally preserved in their
low-dimensional counterparts $y_i$'s and that the redundancy in the
cycles are removed.

To achieve this, a weighted graph $\xi$ is constructed, where each
node corresponds to a cycle $x_i$ and edges are assigned between all
pairs of nodes. The graph can either be fully connected or only bind
those vertices within the $k$-nearest-neighbors and we aims at
extracting the dynamics from the topology of the graph. We use
$W_{ij}$ to denote the similarity between cycle $x_i$ and $x_j$,
which can be chosen conveniently as the correlation coefficient
$\rho_{ij}=Cov(x_i,x_j)/(\sigma_{x_i} \sigma_{x_j})$ (in case $x_i$
and $x_j$ differ in length, shift the shorter vector on the longer
one until $\rho_{ij}$ maximizes). Then, the low-dimensional
representation $y_i$'s can be cast as the solution of the following
optimization problem, $\min \sum_{i,j\in\xi}W_{ij}||y_i-y_j||^2$,
which penalizes those mappings where nearby points $x_i$'s are
relocated far apart in the space of $y_i$'s. In case of univariate
$y_i$'s, the objective can be written as $\textbf{y}^TL\textbf{y}$,
where $\textbf{y} = [y_1, y_2,...,y_k]^{\top}$, $L=D-W$ is the
\emph{graph Laplacian}, and $D$ is the diagonal degree matrix such
that $D_{ii}=\sum_j W_{ji}$. Hence the name \emph{Laplacian
Eiganmap}. To prevent arbitrary scaling in the solution, one can
enforce the constraint $\textbf{y}^TD\textbf{y}=1$.

  \begin{figure}[htbp]
 \begin{center}
 \includegraphics[height=0.3\textwidth]{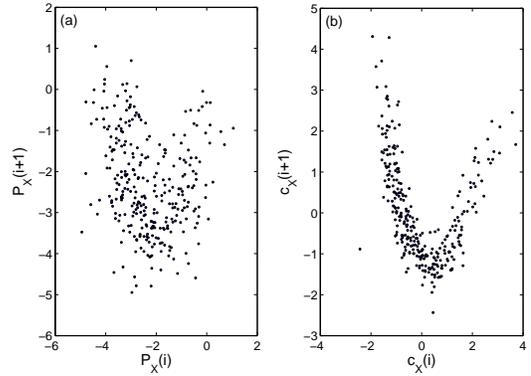}
 \caption{ Return plots for the time series from (a) Poincare section points $P(i)$,
which is obtained by collecting the local minima $X_{min}$, and (b)
the extracted $c(i)$. The R\"{o}ssler system here is corrupted by
$5\%$ dynamical noise and $30\%$ measurement noise. It is obvious
that the return plots from $c(i)$ takes on a clearer form of
quadratic function, indicating that the original nonlinear dynamics
is sufficiently kept.}\label{noise1}
\end{center}
 \end{figure}

The above constrained minimization is solved by the generalized
eigenvalue problem $L \textbf{y}_i=\lambda_i D \textbf{y}_i$, where
$\lambda_i$'s (${i = 1,2,...,k}$) are eigenvalues sorted in an
ascending order, and $\textbf{y}_i$'s are the corresponding
eigenvectors. It can be shown that the minimum eigenvalue
$\lambda_1$ is zero, corresponding to an eigenvector
($\textbf{y}_1$) of all 1's. Therefore it is degenerate and the
optimal solution is actually provided by $\textbf{y}_2$, the
eigenvector of the second smallest eigenvalue \cite{laplacian}. As
is shown in Fig. \ref{ts}, the eigenvector $\textbf{y}_2(i)$
provides a unique, reduced representation of the original time
series by capturing the dynamics of the oscillatory data on the
cycle scale. We use a general notation $c(i)$ ($c$ for cycle) for
such reduced series representation (i.e., other dimension reduction
schemes can also be applied, and we denote the reduced series all as
c(i)). More generally, to obtain an $m$-dimensional solution for
$y_i$'s, one can simply extract the $m$ trailing eigenvectors
$\textbf{y}_{i}$'s ($i=2,3,...,m+1$).

\subsection{The Significance of Extracting Dynamics on Cycle Scale }
The comparison between Laplacian embedding $c(i)$ and the Poincare
section points $P(i)$ (obtained by collecting the local minima)
indicates that they are dynamically identical by sharing the same
chaotic invariants, i.e., Lyapunov exponent and correlation
dimension. However, in case of significant noise (see Fig.
\ref{noise1}), our approach can capture more of the deterministic
structure than does $P(i)$ and is therefore more robust. This is not
surprising, since acquisition of $c(i)$ is based on an optimization
process that preserves the proximity relation between all pairs of
cycles simultaneously, while $P(i)$ is obtained by treating each
cycle independent of each other. It is reasonable to expect the
former to excavate more useful, global patterns than the latter.

\begin{figure}[htbp]
 \begin{center}
 \includegraphics[height=0.3\textwidth]{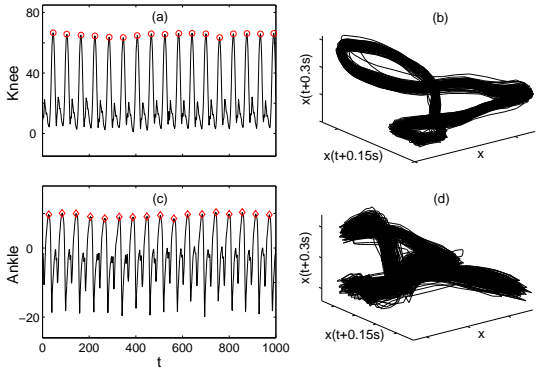}
 \caption{ Time series (left panel) and the corresponding phase space
reconstructions (right panel) for knee ((a)-(b)) and ankle
kinematics data ((c)-(d)). The two time series are typically
non-phase-coherent by demonstrating multi-oscillation within each
cycle. This is also evident from the multi-center rotations of the
attractor in phase space (see (b) and (d)). The two signals are
divided by their respective local maxima into consecutive
cycles.}\label{gait}
\end{center}
 \end{figure}

 Many other forms of dimension reduction can also be applied here,
which produce similar results. For example, the
\emph{Multidimensional Scaling} method that preserves the pairwise
distance yields almost the same outputs as the \emph{Laplacian
Eigenmaps}, see Fig. \ref{ts}(c). Actually, the applicability of
dimension reduction techniques in extracting the dynamics shall be
generally justifiable, considering the intrinsically low correlation
dimensions of most real world pseudoperiodic data. The corresponding
trajectories in phase space tend to have similar orientations for
nearby cycles (see, e.g., Fig. \ref{gait}). Such redundancy can be
effectively removed through dimension reduction, leaving only the
useful degrees of freedom for manifestation of the dynamics of
interest.

 \begin{figure}[htbp]
 \begin{center}
 \includegraphics[height=0.3\textwidth]{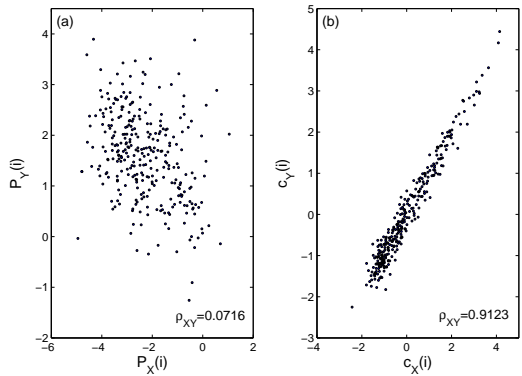}
\caption{Correlation between $X$ and $Y$ component revealed by (a)
Poincare section points $P(i)$ and (b) the extracted $c(i)$. The
R\"{o}ssler system is corrupted by $5\%$ dynamical noise and $30\%$
measurement noise. }\label{noise2}
\end{center}
 \end{figure}

Reconstructing the dynamics on the cycle scale not only enhances the
robustness to noise, but bring new vitality to a number of nonlinear
methods which are otherwise not suitable for oscillatory data due to
the inherent periodicity, such as detrended fluctuation analysis,
surrogate data method, recurrence plot, causality, entropy and so
on. For example, the detrended fluctuation analysis, which computes
the variance of the detrended data at different scales, will produce
similar variance curves (i.e., uprising at small scale and saturate
at the periodicity of the signal) for all pseudoperiodic data
despite their distinct dynamics. This is because the periodic trend
dominating in the data will conceal the underlying, subtle dynamics.
Thus extracting the dynamics on the cycle scale proves to be crucial
for subsequent analysis.

\subsection{Detecting Synchrony from Oscillatory Data} A topic
which is of great interest to oscillatory data analysis in recent
years is to detect the degree of synchronization between
self-sustained oscillators. For example, the peakness of the phase
difference distribution and the consistency of mutual nearest
neighbors are used to characterize the phase synchrony \cite{Tass,
phase} and the dynamical interdependence \cite{a}, respectively.
Here we are especially interested in the case where the phase
relation between two oscillator is evident, but is hard to estimate
due to non-phase-coherence and noise, or is not sensitive to degree
of changes of synchrony. For example, the blood pressure interacted
with heartbeat, where each cycle of blood pressure variation
correspond to one heartbeat. The phase index in this case may not be
quite informative of the synchrony strength. On the other hand, the
presence of noise in most real world data will inevitably degrade
the dependency measures.

 Unlike traditional methods, we
propose to quantify the synchrony the noisy, noncoherent
pseudoperiodic data \emph{on their cycle scale} through their
Laplacian embedding $c(i)$. We first test with the $X$ and $Y$
components of the noisy R\"{o}ssler system. Although these two
components are non-separable and cannot be strictly defined as being
synchronized, they are actually ``in phase'' and therefore suitable
in our context. The two time series are first segmented into cycles
by the local maximums of one of the data (i.e., the \emph{i}th cycle
in both time series share the same time span). We find that the
Poincare section points picked up from the two time series can
hardly capture the interdependency due to noise, in that the scatter
plot between the Poincare section points of $X$ and $Y$ leads to a
group of random points (Fig. \ref{noise2} (a)). Comparatively, the
extracted Laplacian embedding $c(i)$'s extracted from $X$ and $Y$ by
our approach successfully reveal the synchrony pattern by
demonstrating an uprising trend in the corresponding scatter plot
(Fig. \ref{noise2} (b)).

\section{Application to Human Locomotion Data}
\subsection{Characterizing the Locomotion Dynamics}
 Now we apply our approaches to human gait data collected from
electrogoniometers at the knee and ankle joints that measure their
sagittal plane kinematics during overground walking. Human
locomotion is a highly complex, rhythmic process that involves
multilevel control of central nerve system and feedback from various
peripheral sensors. Here we consider two groups of subjects from
healthy controls (CO) and diabetic patients (NP, with significant
peripheral neuropathy), each with 10 subjects \cite{Dingwell}.

\begin{figure}
 \begin{center}
 \includegraphics[height=0.3\textwidth]{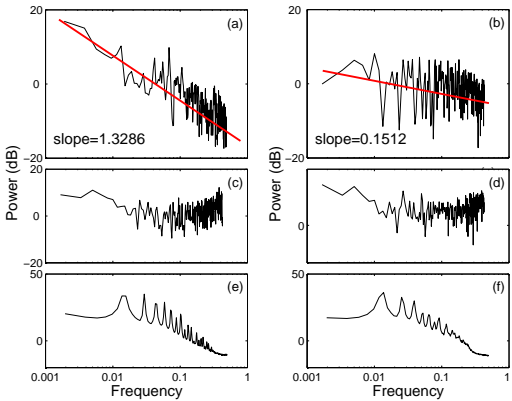}
 \caption{ Power spectrum density (PSD) for typical healthy subject
(the left panel) and the diabetic patient (the right panel). The
top, middle and bottom rows are PSDs for the extracted $c(i)$,
stride interval series, and the original ankle data, respectively.
The frequency here is normalized by the sampling rate. The PSDs for
the stride interval series and the original data show no significant
difference between the healthy subject and the diabetic patient.
}\label{psd}
\end{center}
 \end{figure}

Typically, the human gait time series (Fig. \ref{gait} (a) and (c))
exhibit a relatively fixed frequency with the stride interval
demonstrating little variability, which is suited for dimension
reduction. We first extract the dynamics of the ankle on its cycle
scale by Laplacian embedding. Figure \ref{psd} (a)-(b) display the
power spectrum density (PSD) of the extracted $c(i)$ typically
observed for the two groups. We find that most CO subjects
demonstrate broad band spectrums that scale as $1/f^\beta$,
$\beta=0.76 (mean) \pm 0.23 (std)$, indicating the presence of long
range correlation (i.e., the strides separated by a large time span
are still statistically correlated). In comparison, the spectrum of
the diabetic patients are mostly flat resembling white noise
processes ($\beta=0.37\pm 0.16$), which means that the strides at
different times are mostly uncorrelated. This difference, however,
has not been found with either the stride interval (SI) series (Fig.
\ref{psd} (c)-(d)) or the raw data (Fig. \ref{psd} (e)-(f)). This is
because, the periodicity in the raw data can obscure the underlying
fluctuations and that a linear Fourier transform is not capable of
characterizing nor distinguishing the nonlinear dynamics in the
data; on the other hand, although SI removes the cyclic trend and
has been widely used to quantify pathological locomotion
\cite{Haus}, it loses much of the dynamical information by only
recording the duration of cycles. In comparison, the transform
$c(i)$ successfully removes the periodic trend while preserving the
dynamical fluctuation within each cycle, in particular, the specific
patterns of the four phases within a pace. Therefore it keeps more
valuable information about the neuromuscular control of human
walking.

\subsection{ Synchrony Detection between Knee and Ankle Movement}
The human walking involves the coordination of two major joints,
i.e., the knee and the ankle. So we are also interested in examining
the synchrony between the knee and ankle data, which may provide
further insights in understanding the neuralcontrol of walking. The
knee and ankle movements during continuous walking are obviously in
phase due to the physical connection of the two joints. The strength
of coupling, however, can hardly be recognized from the phase due to
the noncoherence \cite{explain} (see Fig. \ref{gait}). Also, noise
tends to destroy the local structure in phase space and thus hampers
the dynamical dependency measures \cite{qian}. To circumvent these
difficulties, we compare the dynamics of the two time series by
using their Laplacian Embedding $c(i)$'s. Note that each time series
can be segmented by either its own local maximums, or those of its
partner series (markers in Fig. \ref{gait}). Therefore we will
segment each time series twice and compute the averaged correlation
coefficients $\rho_{ij}$'s between $c_{ankle}$ and $c_{knee}$.

 \begin{figure}
 \begin{center}
 \includegraphics[height=0.3\textwidth]{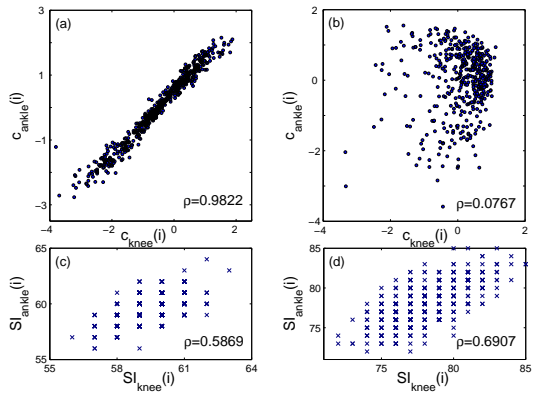}
 \caption{ Synchrony between knee and ankle locomotion
revealed by Laplacian embedding $c(i)$ ((a) and (b)) and stride
interval series ((c) and (d)), with the left and right columns for
an healthy subject and a diabetic patient, respectively. As can be
seen in (b), the stride interval series cannot distinguish the
healthy from the diabetics, as the scatter plots of the stride
interval series show similar uprising trends for both of them.
}\label{y2xianguan}
\end{center}
 \end{figure}

Then we examine the interrelation between the Laplacian embeddings
from the ankle and the knee data. For healthy subjects, the scatter
plots demonstrate a significant uprising trend (Fig.
\ref{y2xianguan} (a), $\rho=0.68\pm0.19$), indicating that the knee
and ankle movements are highly synchronized; while for diabetics
patients, the synchrony almost vanishes (Fig. \ref{y2xianguan} (b),
$\rho=0.26\pm0.18$). Again, the discrimination between CO and NP
groups is unavailable by stride interval series, which always
exhibits a strong correlation between the joints (Fig.
\ref{y2xianguan} (c)-(d)), corresponding to high degree of phase
synchronization. Note that our approach avoids the difficulty of
defining the phase. On the other hand, it is superior to directly
computing the correlation between pairs of equal-time cycles picked
respectively from the two time series, which is sensitive to noise
and nonstationarity. Finally, we point out that a more comprehensive
description of synchrony can be achieved by examining more Laplacian
eigenvectors. In the current case the single eigenvector
$\textbf{y}_2$ already encodes the primary variability and is thus
sufficient for the discriminative task.

Note that we have observed a lack of long range correlation in the
ankle kinematics of the NP group (Fig. \ref{psd} (b)). This suggests
the alteration of the locomotor pattern in the lower limbs, due to
the nerve deterioration in feet and toes typical of the diabetics.
Despite the impaired peripheral feedback caused by the dying nerves,
the \emph{knee} kinematics of the NP group are \emph{still} found to
demonstrate a stable long range correlation indistinguishable from
the CO group. This ``mismatch" between the ankle/knee dynamics for
diabetics can be reasonably explained by the weak synchrony between
these two joints, as is shown in Fig. \ref{y2xianguan} (b). From
these findings, we can see that the impaired neuralfeedback from the
feet influences only the lower limb locomotion while not that of the
knees. We therefore conclude that the human walking is not
critically dependent on the feedback from peripheral nervous system.
It should be noted that although Gates \emph{et.al.} \cite{gates}
checked the same data, they do not consider the interaction among
the knee and angle locomotion which may be important in
characterizing understanding neuralcontrol of human walking.
Furthermore the authors relied solely on the extracted stride
interval series, which have been shown to contain limited dynamical
information and therefore may be insufficient for the analysis and
evaluation of the human walking.

\section{ Conclusion}
To conclude, we have for the first time circumvented the problem of
rebuilding the complex dynamics in oscillatory data from the
viewpoint of dimension reduction. We use a global optimization
procedure to compute the inherent, low-dimensional representation
that maximally preserves the topology of the cycles when embedded in
the multi-dimensional Euclidian space. Our approach has been shown
to be very robust to noise, and is capable of extracting the
underlying dynamics and identifying the subtle synchrony pattern
which usually degrade traditional linear or nonlinear methods.
Application to human gait data provides clinically promising
information in discriminating the healthy from the neuropathological
patients, and further enables us to make fundamental inference on
the neuromuscular control mechanism of human walking.

Our approach may be of great relevance, and is expected to provide
more accurate and powerful diagnostics, to the general biological
and biomedical fields where complex oscillatory data abound. For
example, we have applied the proposed method to the phonation data
from the Parkinson patients, and find a significant lack of long
range correlation among the cycles in the signal compared to the
healthy people. Our approach can therefore serve as an alternative
tool in early Parkinson disease (PD) detection, where there are no
blood or laboratory tests currently that can help in diagnosing PD.
Other potential applications which will be involved in the future
works include the discrimination between Parkinson and essential
tremor, and the evaluation of the interaction between the blood
pressure variation with the heart beat.

\end{document}